\documentclass{appolb}

\usepackage{epsfig}
\usepackage{amsmath}

\newcommand{\be}{\begin{equation}}
\newcommand{\ee}{\end{equation}}
\newcommand{\bea}{\begin{eqnarray}}
\newcommand{\eea}{\end{eqnarray}}
\newcommand{\tr}{{\rm tr}}

\begin{document}
\date{\today}
\pagestyle{plain}
\title{Continuum reduction in large $N$ gauge theories.
\thanks{Lectures at the 49$^{\rm th}$ Cracow School of Theoretical Physics.}}
\author{R. Narayanan
\address{
Department of Physics, Florida International University, Miami,
FL 33199.
\\E-mail: rajamani.narayanan@fiu.edu}}

\maketitle
\begin{abstract}
These are notes associated with three lectures given at
the 49th Cracow School of Theoretical physics where
a pedagogical explanation of
the Gross-Witten transition, Eguchi-Kawai reduction
and continuum reduction were given, followed by a
description of the
numerical computation of 
fermionic observables in the 't Hooft limit
of large $N$ gauge theory.
 
\end{abstract}

\section{Introduction}

It has been a long held hope~\cite{'tHooft:1973jz,'tHooft:1983wm} 
that the large $N$ limit
of QCD is simpler than QCD with $N=3$ but an analytical
solution in $d=4$ is still to come. 
A simplification specific to $d=2$ resulted in an analytical
solution to the large $N$ limit of QCD in $d=2$~\cite{'tHooft:1974hx}.

Relatively recently, one has been able to show that 
continuum reduction~\cite{Narayanan:2003fc} 
holds in $d=4$ and this has been used to make progress in a
numerical solution of large $N$ QCD.
In order to understand continuum reduction,
it will be useful to understand
generalized Eguchi-Kawai reduction~\cite{Eguchi:1982nm}. 
This is best motivated
by studying the large $N$ limit of two-dimensional $U(N)$
lattice gauge theory following Gross and Witten~\cite{Gross:1980he}.

The lectures start off with a pedagogical explanation of the
large $N$ gauge theories on a two dimensional lattice. We will
not include fermions anticipating that they do not play a
dynamical role in $N\to\infty$ limit. After this, the generalized
Eguchi-Kawai reduction will be explained and it is a simple
extension of the original argument by 
Eguchi and Kawai~\cite{Eguchi:1982nm}. 
We will then show that reduction to a single site works only
in $d=2$. Although reduction to a single site does not work
in $d>2$, reduction to a finite physical volume will work
and this is explained in section~\ref{contred}. 

The infinite volume limit at finite $N$ is replaced by a infinite
$N$ limit at finite volume. Spontaneous chiral symmetry breaking
can be realized at finite volume.
Details pertaining
to fermions in the large $N$ limit form the last part of the lectures.

\section{Large $N$ gauge theories in two dimensions\label{gw}}

Consider $U(N)$ gauge theory on an infinite $d$ dimensional
lattice defined by the partition function
\be
Z = \int \prod_x \prod_\mu dU_{x,\mu} e^S;\ \ \
S=bN\sum_p \Tr \left( U_p + U^\dagger_p\right),\label{waction}
\ee
where $S$ is the Wilson action, 
$x$ labels a site, $\mu$ labels a direction and $p$ denotes a 
plaquette. $U_p$ is the parallel transporter around the plaquette.
The inverse 't Hooft coupling is denoted by $b=\frac{1}{g^2N}$. 
Under a  local gauge transformation,
\be
U_{x,\mu} \to g_x U_{x,\mu} g^\dagger_{x+\hat\mu}\label{gtrans}
\ee
and the action is invariant.

We 
follow Gross and Witten~\cite{Gross:1980he}
and gauge fix the two dimensional theory by going to the $A_1=0$ gauge\footnote{This can
be done on a infinite lattice but not on a finite lattice. We will address this
point in section~\ref{ek}.}.
This corresponds to setting $U_{x,1}=1$ for all
$x$. In this gauge,
\be
\sum_p \Tr U_p = \sum_x \Tr \left ( U_{x,2} U_{x+\hat 1,2}^\dagger\right).
\ee 
We still have a remnant gauge symmetry which corresponds to gauge transformations
that are independent of $x_1$, namely $g_{x_2}$,
which can be fixed by setting
\be
U_{x,2}|_{x_1=-\infty} = 1 \forall x_2.
\ee
If we make the change of variables,
\be
U_{x+\hat 1,2}  = U_x U_{x,2},\label{utop}
\ee
then the partition function becomes
\be
Z = \int \prod_x  dU_{x} e^{bN\sum_x \Tr \left( U_x + U^\dagger_x\right)}
= \prod_x 
\int  dU_{x} e^{bN \Tr \left( U_x + U^\dagger_x\right)},\label{raction}
\ee
and it factorizes with plaquettes being independently and identically
distributed. The only remaining symmetry is a global $U(N)$ symmetry,
yet (\ref{raction}) is invariant under
\be
U_x \to V_x U_x V_x^\dagger\label{vtrans}
\ee
for any $V_x$.  
This is special to Yang-Mills action in two dimensions and is not
the original gauge symmetry defined in (\ref{gtrans}). Note that
(\ref{vtrans}) along with (\ref{utop}) implies (\ref{gtrans}) only
if $V_x=g$ for all $x$ and this is the remaining global symmetry.
If we had included fermions, they would have coupled to
$U_{x,2}$ in our gauge and we will not have (\ref{vtrans}) as
a symmetry. 

The factorization of the partition function enables us to compute
expectation values over individual factors and use that result to
get any general expectation value. In particular,
\bea
\langle U_{ij}\rangle &=&
\frac{\int  dU U_{ij} e^{bN \Tr \left( U + U^\dagger\right)}}
{\int  dU e^{bN \Tr \left( U + U^\dagger\right)}}
= \frac{\int  dU dV (VUV^\dagger)_{ij} e^{bN \Tr \left( U + U^\dagger\right)}}
{\int  dU e^{bN \Tr \left( U + U^\dagger\right)}}\cr
&=& \frac{1}{N} \langle \Tr U\rangle \delta_{ij} \equiv w(b,N)\delta_{ij}.\label{eeqn}
\eea
We have used (\ref{vtrans}) in the second equality;
\be
\int dV V_{ij} V^\dagger_{kl} = \frac{1}{N} \delta_{il}\delta_{jk}
\ee
in the third equality and $w(b,N)$ is the expectation value of
a single plaquette.

If we define
\be
z(b,N) = \int  dU e^{bN \Tr \left( U + U^\dagger\right)}\label{sspf}
\ee
as the single plaquette partition function, 
then
\be
w(b,N) = \frac{1}{2N^2} {d\over db} \ln z(b,N).
\ee
The integral in (\ref{sspf}) can be performed~\cite{Bars:1979xb}
 and the result is
\be
z(b,N) = \det M;\ \ \  M_{i,j} = I_{i-j}(2Nb);\ \ \ i,j=1,\cdots,N.
\ee

Consider a rectangular $L\times T$ Wilson loop
with corners at $x$, $x+T\hat 1$, $x+L\hat 2$ and $x+ T\hat 1 + L\hat 2$.
The parallel transporter around this loop in our gauge is
\bea
W_x(L,T) = &&U_{x+T\hat 1,2} U_{x+T\hat 1 + 1\hat 2,2} \cdots 
U_{x+T\hat 1 + (L-1)\hat 2,2} \cr
&&U^\dagger_{x+(L-1)\hat 2 ,2}  
U^\dagger_{x+(L-2)\hat 2,2} \cdots
U^\dagger_{x ,2} . \label{wloop}
\eea
It follows from (\ref{utop}) that
\bea
U_{x+T\hat 1 + (L-1)\hat 2,2} 
U^\dagger_{x+(L-1)\hat 2 ,2} 
= &&U_{x+(T-1)\hat 1 + (L-1)\hat 2} 
U_{x+(T-2)\hat 1 + (L-1)\hat 2}
\cdots \cr
&&U_{x+\hat 1 + (L-1)\hat 2} 
U_{x + (L-1)\hat 2}.
\eea
Using (\ref{eeqn}) and averaging over all the $U_x$ variables
appearing in the above equation, we find that
\be
\frac{1}{N} \langle \Tr W_x(L,T) \rangle = \left[e(b,N\right]^T 
\frac{1}{N} \langle \Tr W_x(L,T-1)\rangle
\ee
Repeating the above steps $L$ times we arrive at 
\be
\frac{1}{N} \Tr W_x(L,T) = \left [ e(b,N)\right]^{LT},\label{arealaw}
\ee
since $W_x(L,0) = 1$.
The above equation says that the area law is exact in two
dimensional Yang-Mills theory for all values of $N$.

The continuum limit at a fixed $N$ is obtained by taking $b\to \infty$.
If we first take $N\to \infty$ at a fixed $b$ and then take $b\to \infty$,
we obtain the continuum limit of the large $N$ gauge theory \`a la 't Hooft.
In the large $N$ limit, it is instructive to solve for $z(b,N)$ using the method
of steepest descent resulting in the stationary condition
\be
2b\sin\alpha_i = \frac{1}{N} \sum_{j\ne i} \cot \left| \frac{\alpha_i - \alpha_j}{2}\right|
\label{aleqn}
\ee
for the eigenvalues $\alpha_i$ of the $U(N)$ matrix, $U$, appearing in (\ref{sspf}).
Due to the symmetry (\ref{vtrans}) of the single site partition function,
all expectation values will be only functions of the eigenvalues of $U$
and can be evaluated by substituting for  $\alpha_i$ the values that solve
(\ref{aleqn}). 
Since the single site partition function is dominated by the
stationary point, it follows that expectation values factorize in the large $N$
limit:
\be
\langle F(U) \rangle = F (\langle U \rangle).\label{foper}
\ee
The partition function on the infinite lattice can be further reduced from a
product of independently and identically distributed plaquettes in (\ref{raction})
to a single site partition function as in (\ref{sspf}). 
Expectation of the Wilson loop operator defined in (\ref{wloop}) reduces
to the {\sl folded} operator
\be
W_x(L,T) =\langle  Tr U^{LT} \rangle,
\ee
and it is clear from (\ref{foper}) that we will obtain (\ref{arealaw}).
This is the motivation behind the idea of Eguchi-Kawai reduction~\cite{Eguchi:1982nm}.

Before we proceed with a discussion of Eguchi-Kawai reduction, it will be
useful to finish this section with a property of (\ref{aleqn}).
We can replace, $\alpha_i$, by a continuum function
$\alpha(x)$, $x\in [0,1]$  in the large $N$ limit. Furthermore, we can
define the density of eigenvalues,
\be
\rho(\alpha) = \frac{dx}{d\alpha}.
\ee
Then (\ref{aleqn}) reduces to 
\be
2b\sin\alpha = P \int_{-\pi}^{\pi} d\beta \rho(\beta)
\cot \frac{\alpha - \beta}{2},
\ee
an equation for $\rho$ 
where $P$ refers to the principal part of the integral.
This equation is solved in~\cite{Gross:1980he} and the result is
\be
\rho(\alpha) = 
\begin{cases}
\frac{2b}{\pi} \cos\frac{\alpha}{2}\sqrt{\frac{1}{2b} - \sin^2\frac{\alpha}{2}}
& \text{if $b \ge \frac{1}{2}$ and $|\alpha| < 2 \sin^{-1}\sqrt{\frac{1}{2b}}$} \cr
\frac{1}{2\pi}\left( 1 + 2b\cos\alpha\right) &
\text{if $ b \le \frac{1}{2}$ and $|\alpha| \le \pi$}\cr 
\end{cases}
\ee
The lattice theory undergoes a phase transition at $b=\frac{1}{2}$.
The continuum theory does not exhibit this phase transition. But
the lattice strong coupling limit and the weak coupling limit
are separated by this phase transition. In order to obtain the
correct continuum limit of the large $N$ theory, we need to keep
$b > \frac{1}{2}$ and $\rho(\alpha)$ has a finite region of
support around $\alpha=0$ that does not extend up to 
$\alpha=\pm\pi$. 

\section{Generalized Eguchi-Kawai reduction\label{gek}}

The discussion of the large $N$ limit of Yang-Mills theories in
section~\ref{gw} suggests:
\begin{enumerate}
\item Factorization of observables;
\item Domination of the path integral by a single {\sl classical} configuration.
\end{enumerate}
Witten argues for the above two points in~\cite{Witten:1979pi}.
Consider, for example, an observable that we encountered in 
section~\ref{gw}, namely, $\langle \Tr U_x \Tr U_y \rangle$.
This quantity, in perturbation theory, has two pieces, connected and disconnected.
$\langle Tr U_x\rangle$ is of order $N^2$ since there are $N^2$ gluon species
that can run around the loop. Therefore, the disconnected piece is of order $N^4$.
The connected piece has only one loop with two insertions, one for $\Tr U_x$
and another for $\Tr U_y$, and therefore it is of order $N^2$.  Therefore,
\be
\langle \Tr U_x \Tr U_y \rangle = \langle \Tr U_x\rangle \langle \Tr U_y\rangle
\ee
in the large $N$ limit. The same argument would also imply that
\be
< e^2> - <e>^2 =0;\ \ \ \  e = \frac{1}{V} \sum_x \Tr U_x.
\ee
If fluctuations go to zero in the large $N$ limit, it is not necessary to do a path
integral since one classical field configuration must dominate as was
seen using steepest descent in section~\ref{gw}. 
The above argument of Witten was made rigorous in~\cite{Eguchi:1982nm}
and  the following statement is a generalization of the Eguchi-Kawai reduction:

Consider $U(N)$ Yang-Mills gauge theory with Wilson action given by
(\ref{waction}) on
a finite lattice of size $L_1\times L_2\cdots L_d$ with a fixed lattice
coupling $b=\frac{1}{g^2N}$ and periodic boundary conditions in all
directions. 
Also consider another theory with only one difference from the 
previous one: $L_{\mu_i}=\infty$ for $i=1,\cdots, k \le d$.
Now consider an arbitrary closed Wilson loop operator.  The operators
associated with the same Wilson loop on the finite lattice and
the lattice with $k$ infinite directions could be different due
to possible folding.
The folding comes from the use of periodic boundary
conditions on the finite lattice. 
The large $N$ limit will be the same in both cases
provided the $Z_N$ symmetries associated with
the Polyakov loops in the $\mu_i$; $i=1,\cdots,k$ directions are not broken
on the finite lattice.

We provide relevant steps for
a proof of the above statement by following the steps in~\cite{Eguchi:1982nm}.
Consider a closed Wilson loop that contains the link $U_{x,\mu}$ once
and let us write it is as 
\be
W = \Tr U_{x,\mu} C^\dagger_{x,\mu}
\ee
where
$C_{x,\mu}$ is an open path with more than one link
that connects $x$ and $x+\hat\mu$ and does not contain
$U_{x,\mu}$.
The terms in the Wilson action that contain $U_{x,\mu}$ can be
written as 
$\Tr \left ( U_{x,\mu} S^\dagger_{x,\mu} +
S_{x,\mu} U^\dagger_{x,\mu}\right)$  
where $S_{x,\mu}$ is the sum 
of the parallel transporters over
all the three link paths that connect $x$ and $x+\hat\mu$.
$S_{x,\mu}$ does not contain $U_{x,\mu}$ if none of the finite
directions are of unit length\footnote{Equation (\ref{sdeqn}) 
remains unaltered if $S_{x,\mu}$ contains $U_{x,\mu}$.}.
The group measure is invariant under a small change of the form
\be
U_{x,\mu} \to e^{i\epsilon T^j} U_{x,\mu}
\ee
where $T^j$ is a group generator and $\epsilon$ is a small parameter.
Therefore,
\bea
\langle \Tr \left(T^j U_{x,\mu} C^\dagger_{x,\mu} \right)\rangle
= 
\langle && \Tr \left( T^j e^{i\epsilon T^j} U_{x,\mu} C^\dagger_{x,\mu}\right) \cr
&& e^{bN\Tr \left ( \left[e^{i\epsilon T^j} -1\right]U_{x,\mu} S^\dagger_{x,\mu}
+
S_{x,\mu} U^\dagger_{x,\mu}]\left[e^{-i\epsilon T^j} -1\right]\right)}
\rangle,
\eea
and to the lowest order in $\epsilon$,
\bea
&&\langle \Tr \left(T^j T^jU_{x,\mu} C^\dagger_{x,\mu} \right)\rangle
\cr 
&&+bN
\langle \Tr \left(T^j U_{x,\mu} C^\dagger_{x,\mu} \right)
\Tr\left(T^j U_{x,\mu}S^\dagger_{x,\mu}
- S_{x,\mu}U^\dagger_{x,\mu}T^j\right)\rangle = 0
\eea 
Summing the above equation over all values of $j$ and using
the identity,
\be
\sum_{j=1}^{N^2} T^j_{ab} T^j_{cd} = \delta_{ad}\delta_{bc},
\ee
we get
\be
\langle \Tr \left(U_{x,\mu} C^\dagger_{x,\mu} \right)\rangle
+b
\langle \Tr \left(U_{x,\mu} C^\dagger_{x,\mu} U_{x,\mu}S^\dagger_{x,\mu}
\right)
\rangle
- b
\langle \Tr \left(C^\dagger_{x,\mu} S_{x,\mu}\right)\rangle = 0
\label{sdeqn}.
\ee
The above equation, referred to as the
Schwinger-Dyson equation for Wilson loops,
relates the expectation value of the original
Wilson loop to other Wilson loops that correspond to the
modification of the original Wilson loop by attaching the various
plaquette parallel transporters that contain $U_{x,\mu}$.
Even if we start with a Wilson loop where each link occurs
only once, the above equations will generate Wilson loops
where certain links appear more that once. In fact, this can
already been seen in the second term of (\ref{sdeqn}). If we
repeat the above procedure starting with a Wilson loop
where $U_{x,\mu}$ appears twice, namely,
\be 
W = Tr U_{x,\mu} C^\dagger_{x,\mu} U_{x,\mu} D^\dagger_{x,\mu}
\ee 
where $C_{x,\mu}$ and $D_{x,\mu}$ are two open paths that connect $x$ to
$x+\hat\mu$. In this case, the Schwinger-Dyson equation will have
an additional term of the form
\be
\frac{1}{N}
\langle 
\Tr \left(U_{x,\mu} C^\dagger_{x,\mu} \right)
\Tr \left(U_{x,\mu} D^\dagger_{x,\mu} \right)
\rangle.
\ee
This is an expectation value of products of Wilson loops.
Noting that expectation value of an closed loop is of order $N$,
we see that the new term is the same order in $N$ as the ones in
(\ref{sdeqn}) and we also note that the new term factorizes in
the large $N$ limit. 
The coupled set of infinite number of Schwinger-Dyson equations
obtained in this process will also involve Polyakov loops in
the finite directions
for the following reason:
Since it will involve Wilson loops of arbitrary size, it will contain
loops of the form
\be
W = Tr U_{x,\mu} C^\dagger_{x,\mu} 
U_{x+L_{\mu}\hat\mu,\mu} D^\dagger_{x,\mu}
\ee
where
$C_{x,\mu}$ is an open path that connects $(x+L_{\mu}\hat\mu)$ to $x+\hat\mu$
and $D_{x,\mu}$ is an open path the connects
$x$ to $(x+(L_{\mu}+1)\hat\mu)$.
Since $U_{x+L_\mu\hat\mu,\mu}=U_{x\mu}$ by periodic boundary conditions,
this will result in a term of the form
\be
\frac{1}{N}
\langle 
\Tr \left(U_{x,\mu} C^\dagger_{x,\mu} \right)
\Tr \left(U_{x+L_\mu\hat\mu,\mu} D^\dagger_{x,\mu} \right)
\rangle.\label{ploopterm}
\ee
The difference between the infinite set of coupled equations in the
two cases, one with $k$ infinite directions and the other being finite
in all directions, is the presence of additional Polyakov loops in the $k$
finite directions. Polyakov loops in the $\mu$ direction are not
invariant under a global $Z_N$ symmetry in that direction where we replace
all $U_{x,\mu}$ in a fixed hyperplane perpendicular to $\mu$ by
$e^{i\frac{2\pi k}{N}} U_{x,\mu}$ with $0 < k < N$.
Since this is a symmetry of the gauge action,
the Polyakov loops appearing in (\ref{ploopterm})
will have zero expectation value if the $Z_N$ symmetry in the
$\mu$ direction is not spontaneously broken. This completes
our discussion of the statement concerning generalized Eguchi-Kawai reduction.

\section{Reduction to a single site\label{ek}}

The arguments presented in section~\ref{gek} show that one can
reduce the large $N$ theory from an infinite lattice down to a single
site lattice if the $Z_N$ symmetries on a single site lattice are
not broken. Since we independently showed that this reduction
was possible in two dimensions in section~\ref{gw}, it follows
that the two $Z_N$ symmetries are not broken on a single
site lattice in two dimensions for all values of $b$ and therefore
also in the continuum limit.  On the other hand, the $Z_N$ symmetries
are broken in the weak coupling limit in three or more dimensions
and we will present the argument following~\cite{Bhanot:1982sh}.

Consider the $U(N)$ Wilson gauge action on a single site $d$ dimensional lattice, namely,
\be
S_{\rm EK} = bN \sum_{\mu \ne \nu=1}^d \Tr \left[ U_\mu U_\nu U_\mu^\dagger
U_\nu^\dagger\right].
\ee
The action depends on $d$ $U(N)$ matrices and the gauge transformation is
\be
U_\mu \to g U_\mu g^\dagger.
\ee
Note that the eigenvalues of $U_\mu$ are gauge invariant.
We cannot fix a gauge such that one of the $U_\mu=1$ since we are on a finite
lattice. The action has an additional $U^d(1)$ symmetry given by
\be
U_\mu \to e^{i\phi_\mu} U_\mu;\label{znsymm}
\ee
with $ -\pi < \phi_\mu < \pi$. 
The four Polyakov loop operators given by
\be
P_\mu = \Tr U_\mu 
\ee
are gauge invariant but not invariant under (\ref{znsymm}). 
If the $U^d(1)$ symmetry is not broken, then the eigenvalues of
all $U_\mu$ are uniformly distributed on the unit circle and $P_\mu=0$.
In order to see if this symmetry is spontaneously broken
in the weak coupling limit, we set
\be
U_\mu = e^{ia_\mu} D_\mu e^{-ia_\mu};\ \ \
D_\mu^{jk} = e^{i\theta^j_\mu}\delta^{jk},
\ee
and expand to the quadratic term in the hermitian matrix $a_\mu$.
We fix the gauge by setting $a_1=0$. 
The group measure is given by
\be
\prod_\mu dU_\mu = \left[ \prod_\mu 
\prod_i d\theta^i_\mu \right] \left[\prod_\mu \prod_{i > j} p^{ij}_\mu \right] 
\left[\prod_{\mu=2}^d
\prod_{i>j} 
da^{ij}_\mu
d{a^{ij}_\mu}^*\right]
\ee
where
\be
p^{ij}_\mu = \sin^2 \frac{1}{2}\left (\theta^i_\mu - \theta^j_\mu\right).
\ee
The quadratic piece of the action is
\be
S = - 32 bN \sum_{i < j} \sum_{\mu, \nu=2}^d  {a_\mu^{ij} }^* 
\left ( p_\mu^{ij} p^{ij} \delta_{\mu\nu} - p_\mu^{ij} p_\nu^{ij} \right)
a_\nu^{ij},
\ee
where
$p^{ij} = \sum_\mu p_\mu^{ij}$.
The result of the integration over $a_\mu$, ignoring normalization factors,
is
\be
\left[ \prod_\mu \prod_{i>j} \frac{1}{p_\mu^{ij}}\right]   \left[ \prod_{i>j}  p^{ij}\right]^{2-d}
\ee
Therefore, up to second order in $a_\mu$, the partition function is
\be
Z =  \left[ \prod_\mu 
\prod_i d\theta^i_\mu \right] 
e^{(2-d)\sum_{i<j} \ln p^{ij}}.\label{zquad}
\ee
All $\theta_\mu^i=0$ has the maximum probability for $d>2$ since
$p^{ij}=0$
for all $i$ and $j$ implying that the $Z_N$ symmetries are broken in
the
weak coupling limit if $d>2$.

\section{Continuum reduction \label{contred}}

Consider the continuum theory in a finite torus of
size $l_1\times l_2\cdots l_d$ obtained from the theory on a 
$L_1\times L_2\cdots L_d$ periodic lattice at a fixed coupling $b$ and 
taking the limit $L_1,L_2,\cdots,L_d,b\to \infty$ such that 
\bea
l_i &=& \frac{ L_i}{\sqrt{b}};\ \ \  i=1,2, \ \ \ {\rm in}\ \  d=2 \cr
l_i &=& \frac{ L_i}{b};\ \ \  i=1,2,3 \ \ \ {\rm in}\ \  d=3 \cr
l_i &=& L_i a(b);\ \ \ i=1,2,3,4\ \ \ {\rm in}\ \ d=4
\eea
are kept fixed. The lattice spacing, $a(b)$, in $d=4$ is given by
\be
a(b) = \frac{1}{\Lambda} \left(\frac{48\pi^2 b}{11}\right)^{\frac{51}{121}}
e^{-\frac{24\pi^2 b}{11}},
\ee
in weak coupling perturbation theory~\cite{Creutz:1984mg}.
 
In two dimensions, the variables, $\theta_\mu^i$ are uniformly
distributed in the weak coupling limit as can be seen by setting $d=2$
in (\ref{zquad}).  Therefore the single site lattice $L_1=L_2=1$ will
give the same results in the large $N$ limit as any $L_1\times L_2$
lattice for all values of the coupling $b$ implying that the continuum
theory will be independent of $l_1$ and $l_2$. Since the $Z_N$ ($U(1)$
in the limit $N\to\infty$) symmetries are unbroken for all $l_1$ and
$l_2$, it follows that two dimensional large $N$ QCD is in the confined phase for all
temperatures.

The lack of a finite temperature phase transition separating the
confined
phase from the deconfined phase is special to $d=2$. Numerical
analysis has strongly established the existence of the deconfining
phase transition 
for finite $N$ in 
$d=3$~\cite{Engels:1985tz,Gross:1984pq} and 
$d=4$~\cite{McLerran:1980pk,Iwasaki:1992ik,Gavai:2002td}. 
This transition is expected to
have a large $N$ limit both
in $d=3$~\cite{Liddle:2005qb} 
and $d=4$~\cite{Lucini:2003zr}.
Therefore, we could not have expected single site reduction to work in
$d>2$. 

Consider a symmetric lattice $L_1=L_2=\cdots L_d=L$ ($d>2$) and a lattice
coupling $b_1(L)$ such that no $Z_N$ symmetry is broken for $b< b_1(L)$
but not all $Z_N$ symmetries are unbroken for $b> b_1(L)$.
We know that such a coupling exists since all $Z_N$ symmetries will be
unbroken for $b=0$ and all will be broken for $b=\infty$.
The theory is in the confined phase for $b < b_1(L)$ and since
we expect the continuum theory to have a confined phase $b_1(L)$
should approach $\infty$ as $L\to\infty$. Fixing, $b< b_1(L)$ we can
consider a $L_1\times\L_2\cdots L_d$ lattice such that $L_i > L$ for
all $i$ and the arguments provided in section~\ref{gek} shows that
there will be no dependence on $L_i$ and the theory is in the confined phase.
Now consider a $L\times
\infty^{d-1}$ lattice with a coupling of $b=b_1(L)$. As $L$ is varied
we will remain at the phase transition point separating the confined
phase from the deconfined phase where the $Z_N$ symmetry is broken in the
direction with finite extent $L$.  Therefore, $l_1=L/b_1(L)$ in $d=3$
and $l_1=La(b_1(L))$ in $d=4$ should have a finite limit as
$L\to\infty$ and it should be the inverse of the deconfining
temperature. This has been numerically verified in 
$d=3$~\cite{Narayanan:2003fc} and $d=4$~\cite{Kiskis:2003rd}.
The order of the deconfining phase transition at infinite $N$ can be
obtained by a numerical computation of the latent heat associated
with the transition. The presence of a latent heat will result in a
jump in the average value of the action density at the transition.
A non-zero latent heat has been computed in $d=4$~\cite{Kiskis:2005hf} by numerical
methods and numerical studies are currently under way in $d=3$~\cite{koren}.

We will refer to the confined phase of the continuum theory
as the $0$c phase. There is no dependence on the physical size of
the box in this phase. There is a transition from the $0$c to
the $1$c phase when one of the directions has a length less than
$l_1$. This is the conventional deconfined phase and there is no
dependence on the physical size of the box in the $(d-1)$ directions.
Let $l< l_1$ be the length of the direction along which the $U(1)$
symmetry is broken. Let $(d-2)$ of the other directions be finite
and let one direction be finite. Fixing $l$, we can vary the length
of the second finite direction and go from a $U(1)$ symmetric
phase in that direction to a $U(1)$ broken phase. Let $l_2(l) \ge l $ be
the length of the second finite direction such the the U(1) symmetry
in that direction is unbroken for lengths larger than $l_2(l)$ and
broken for lengths smaller than $l_2(l)$. The existence of $l_2(l)$
in the continuum theory has been verified by numerical means using
the lattice theory for $d=3$~\cite{Narayanan:2007ug}. The temperature in the
deconfined phase is $1/l$ and one can remain in the deconfined
phase for all temperatures above $1/l_1$ as long as one keeps
the extent of the other $(d-1)$ directions larger than $l_2(l)$.
If we pick one of the other $(d-1)$ directions to be less than $l_2(l)$
two of the $U(1)$ symmetries are broken and we refer to this as
the $2$c phase. In this manner we can have the continuum theory
in a $k$c phase with $0\le k \le d$ where $k$ of the $d$ $U(1)$ symmetries
are broken. One can view the $(d-1)$c and the $d$c phase as the
low temperature and the high temperature phase of large $N$ QCD
in a Bjorken universe~\cite{Bjorken:1979hv}. The complete phase
diagram has not been mapped out in $d=3$ or $d=4$.

\section{Fermions}

As long as the fermions are in the fundamental representation and
we only have a finite number of flavors, $N_f$, fermion loops are suppressed
in the large $N$ limit compared to gluon loops since $NN_f << N^2$
as $N$ gets large~\cite{'tHooft:1973jz, 'tHooft:1983wm}. 
Physical quantities associated with the fermionic sector of large $N$
QCD can be computed using fermionic observables in a gauge background
generated using the pure gluonic action. Continuum reduction continues
to hold while computing physical quantities in the fermionic sector.

\subsection{Chiral condensate}

Chiral symmetry is expected to be broken in $d=2$ and $d=4$ in the 
confined phase of large
$N$ limit of QCD. The theory in $d=2$ is in the confined phase
for any finite torus. The large $N$ degrees of freedom must
therefore be responsible for spontaneous chiral symmetry breaking
in finite volume. Consider the lattice model on a single site.
We have two $U(N)$ matrices, namely, $U_1$ and $U_2$. The Wilson
action in (\ref{waction}) reduces to $S=2bN {\rm Re\ } \Tr \left[
U_1 U_2 U_1^\dagger U_2^\dagger\right]$.
The fermionic operator on the infinite lattice splits into momentum
blocks with each block being of the form $D_f\left(U_1e^{ip_1},U_2e^{ip_2};m_q\right)$ where $-\pi < p_1,p_2 \le \pi$
is the momentum of the block and $m_q$ is the
quark mass. The chiral condensate is given by
\be
\chi(b,N,m_q) = \frac{1}{(2\pi)^2} \int_{-\pi}^\pi
dp_1 \int_{-\pi}^\pi dp_2 
\frac{1}{N} \langle \Tr D_f\left(U_1e^{ip_1},U_2e^{ip_2};m_q\right)\rangle,
\ee
where the expectation value is obtained using the single site Wilson
gauge action. In order to obtain the chiral condensate, we will have
to take $N\to\infty$ limit before we take the $m_q\to 0$ limit.
In the limit of $N\to\infty$, the two $U(1)$ symmetries are not
broken and $\langle \Tr D_f\left(U_1e^{ip_1},U_2e^{ip_2};m_q\right)\rangle$ is independent of $p_1$ and $p_2$. 
Therefore, the chiral condensate in the massless limit is given by
\be
\Sigma = \lim_{m_q\to 0}
\lim_{N\to\infty}
\frac{1}{N} \langle \Tr D_f\left(U_1,U_2;m_q\right)\rangle.
\ee

The fermionic operator $D_f(U_1,u_2;0)$ on the single
site lattice will have $2N$ eigenvalues
and it will have $N$ paired eigenvalues, $\pm\lambda_i$; 
$i=1,\cdots, N$ if the global topology is zero and the fermionic
operator obeys chiral symmetry on the lattice.
Let us assume that $0 < \lambda_1 < \lambda_2 < \cdots < \lambda_N$
and let $p(\lambda_1,\lambda_2,\cdots,\lambda_N)$ be
the joint probability  of the $N$ eigenvalues after averaging over
$U_1$ and $U_2$ using the single site gauge action as the measure.
Chiral random matrix theory~\cite{Shuryak:1992pi} predicts the joint probability
distribution, $p_{\rm chRMT}(z_1,z_2,\cdots,z_N)$, when
$N$ is large and where
\be
z_i = \Sigma(b) N\lambda_i
\ee
with $\Sigma(b)$ being the only adjustable parameter and is
the chiral condensate at the lattice coupling $b$. This
prediction has been numerically verified~\cite{Narayanan:2004cp} and the resulting
chiral condensate agrees 
with the known analytical result~\cite{Burkardt:2002yf,Zhitnitsky:1985um}. 

Turning now to $d=4$, we first note following the discussion
in section~\ref{contred} that we need to
consider a theory on
a $L^4$ lattice at a fixed lattice coupling $b$ 
such that $b < b_1(L)$ and one is in the $Z_N$ symmetric phase. 
The fermionic operators on the $L^4$ lattice will be such that
\be
\psi(x+L\hat\mu) = e^{ip_\mu}\psi(x)\label{ferbc}
\ee
since the operator in the infinite lattice will split into momentum
blocks with the momentum translating into the above boundary
conditions on fermions. We can replace the boundary conditions
on fermions by periodic boundary conditions and replace the
gauge fields by
\be
U_\mu(x) \to U_\mu(x) e^{i\frac{p_\mu}{L}},
\ee
where we have used a gauge transformation to uniformly distribute
$e^{ip_\mu}$ over $L$ links.
Since we are in the $U(1)$ symmetric phase of the large $N$ gauge
theory, the chiral condensate will not depend on $p_\mu$
and we can set them to zero and compute the chiral condensate.
One can use matching with chiral random matrix theory predictions
to numerically estimate the chiral condensate in the large $N$ limit
of QCD in $d=4$. 
The eigenvalues, $\pm\lambda_i$, $i=1,\cdots,4NL^4$, will
match with the variables, $z_i$, according to
\be
z_i = \Sigma(b) N L^4 \lambda_i.
\ee
The chiral condensate, $\Sigma(b)$, has been numerically extracted
and its scaling behavior has been studied as a function of $b$~\cite{Narayanan:2004cp}.

The fermion propagator in the deconfined ($1$c) phase will depend on
the boundary conditions, namely (\ref{ferbc}), in the direction where
the $Z_N$ symmetry is broken. Fermions, in this sense, do play
a dynamical role in the $1$c phase and anti-periodic boundary
conditions with
respect to the value of the Polyakov loop in that direction will
be favored. This has been numerically verified
in~\cite{Narayanan:2006sd}.
Since the chiral condensate does not depend on the temperature in
the confined phase, the chiral phase transition in going from $0$c
to $1$c will be first order. Restoration of chiral symmetry in the
$1$c phase has been numerically verified by the presence of
a non-zero gap in the fermion spectrum for all values of temperature
in the deconfined phase~\cite{Narayanan:2006sd}.

\subsection{Meson propagator}

Let
\be
M(x) = \bar u(x) \Gamma \sum_z S_{xz}(U_\mu) d(z)
\ee
denote a meson at $x$ made out of two different flavors in some
spin representation given by
$\Gamma$. The meson is defined using some gauge field dependent
smearing operator, $S_{xy}(U_\mu)$, that commutes with $\Gamma$
and transforms as
\be
S_{xz}(U^g_\mu) = g_x S_{xz}(U_\mu) g^\dagger_z
\ee
under a gauge transformation, $d(z) \to g_z d(z)$ and $\bar u(x) \to
\bar u(x) g^\dagger_x $.
The meson propagator in momentum space is
\bea
G(p) &=& \sum_{x,y} e^{ip(x-y)} \langle M(x) M^\dagger(y) \rangle\cr
&=& \int \frac{d^4q}{16\pi^4}
\sum_{x,y,z,w}\Bigg\langle \tr 
\Biggl[ \Gamma  e^{i(\frac{p}{2}+q)x}
S_{xz}(U_\mu)G^f_{zw}(U_\mu,m_q)
S^\dagger_{wy}(U_\mu)
e^{-i(\frac{p}{2}+q)y} 
\cr
&&\ \ \ \ \ \ \ \ \ \ \ \ \ \ \ \ \ \ 
\Gamma^\dagger  e^{-i(\frac{p}{2}-q)y}  
G^f_{yw} (U_\mu,m_q) e^{i(\frac{p}{2}-q)x} 
\Biggr]\Bigg\rangle\cr
&=& \int \frac{d^4q}{16\pi^4}
\Bigg\langle \Tr \Biggl[ \Gamma S(U_\mu e^{i(\frac{p}{2}+q)_\mu})
G^f(U_\mu e^{i(\frac{p}{2}+q)_\mu},m_q) S^\dagger (U_\mu
e^{i(\frac{p}{2}+q)_\mu}) \cr
&&\ \ \ \ \ \ \ \ \ \ \ \ \ \ \ \ \ \
\Gamma^\dagger G^f(U_\mu e^{i(-\frac{p}{2}+q)_\mu},m_q) 
\Biggr]\Bigg\rangle\cr
&=& \Bigg\langle \Tr \Biggl[ \Gamma S(U_\mu e^{i\frac{p_\mu}{2}})
G^f(U_\mu e^{i\frac{p_\mu}{2}},m_q) S^\dagger (U_\mu e^{i\frac{p_\mu}{2}})
\Gamma^\dagger G^f(U_\mu e^{-i\frac{p_\mu}{2}},m_q) 
\Biggr]\Bigg\rangle\cr
&&
\eea
The first equality above assumes that we have translational invariance
upon averaging over the gauge fields. We have introduced an integral
over $q$ in the second equality and the integrand does not depend
upon $q$. The $\tr$ in the second equality indicates a sum over spin
and color indices only. We extend this to $\Tr$ in the third equality
where the sum is now over space, spin and color indices.
The exponential factors on either side of the smeared $d$ quark
propagator and on either side of the $u$ quark propagator
in the second equality
are viewed as a gauge transformations and results in the
gauge transformed fields in the third equality. These factors
are thought of as momenta carried by the quarks: $d$ quark
has a momentum equal to $q+\frac{p}{2}$ and the $u$ quark
has a momentum equal to $q-\frac{p}{2}$. This corresponds to
a meson momentum equal to $p$ and $q$ is the momentum around
the quark loop in the meson propagator.
Due to the $U(1)$ gauge invariance in the confined phase,
we can replace $U_\mu e^{iq_\mu}$ by $U_\mu$ in the third equality
and the integrand does not depend on $q$. This results
in the final equality in the above equation.

The meson momentum can take any value in the range $[-\pi,\pi]$ since
the above equation was derived on an infinite lattice. Momenta that
are integer multiples of $\frac{2\pi}{L}$ are the ones allowed by 
periodic boundary conditions on the $L^4$ lattice. Momenta
that fill in the gaps between the integer multiples correspond to
boundary conditions of the form given by (\ref{ferbc}).

Pion mass and the vector meson mass as a function of the quark
mass has been numerically studied using the above procedure.
This has resulted in an estimate of the pion decay constant~\cite{Narayanan:2005gh}
and the vector meson mass in the chiral limit~\cite{Hietanen:2009tu}.

\subsection{Numerical details}

We will focus on $d=4$ and present the necessary numerical details to
perform the computations described in the previous subsections.
Gauge fields
are generated using a combination of Cabibbo-Marinari
$SU(2)$ heat-bath~\cite{Cabibbo:1982zn} and $SU(N)$
over-relaxation~\cite{Kiskis:2003rd}.
A description of overlap
fermionscan be found
in~\cite{Neuberger:1997fp,Narayanan:1994gw,Edwards:1998wx}.
The $\epsilon$ function appearing in the overlap-Dirac
operator is best approximated using the $20^{\rm th}$ order 
Zolotarev approximation~\cite{vandenEshof:2002ms,Chiu:2002eh}.
The action of $\epsilon$ on a vector can be performed using the
multiple mass conjugate gradient algorithm~\cite{Jegerlehner:1996pm,Jegerlehner:1997rn}.
The low lying eigenvalues of the massless overlap-Dirac can be computed
using the Ritz algorithm~\cite{Kalkreuter:1995mm}.
The trace involved in the computation of the meson propagator can be
stochastically estimated using a single random vector.

\section{Other topics and future directions}

Some progress in the understanding of the physical transition from 
strong coupling (hadron resonances) to
weak coupling (perturbative QCD) has been achieved by the study
of Wilson loops~\cite{Neuberger:2009wm}. 
Much more work needs to be done in this topic and there is recent
progress in connecting the transition in Wilson loops to a chiral
transition in two-dimensional fermions coupled to the four
dimensional gauge field~\cite{Narayanan:2009ag}. 

A computation of the string tension using folded Wilson loops
has been performed in $d=3$~\cite{Kiskis:2008ah} and
$d=4$~\cite{Kiskis:2009rf}.
Typically, one obtains the string tension from Polyakov loop
correlators~\cite{Bringoltz:2006zg} since large Wilson loops have a small expectation
value due to two reasons: large area and perimeter divergence.
The results at large $N$ obtained using folded Wilson loops are in
good agreement with the results obtained at smaller $N$ using
Polyakov loop correlations and then performing a large $N$
extrapolation~\cite{Bringoltz:2006zg,Lucini:2005vg}.
Based on our discussion in section~\ref{gek}, we note that
folded Wilson loops can also be computed in the $1$c phase
as long as the Wilson loop is in the plane where the $Z_N$ symmetries
are not broken and such loops are referred to as spatial loops. 
Numerical computation of the spatial string tension in the
$1$c phase of $d=3$ show that the string tension grows linearly
with the temperature~\cite{Kiskis:2009xj}. 

Fermions play a dynamical role in the large $N$ limit when they
are in the adjoint representation. Since the gauge action
induced by the fermion determinant
will tend to cancel the original gauge action, one expects single site
reduction to hold in $d=4$ if the fermions are in the adjoint
representation. Strong arguments in this direction have been
put forward recently in
the continuum~\cite{Kovtun:2007py,Unsal:2008ch}. 
Taking the zero volume limit in the continuum
is not necessarily the same as working on a single site lattice.
Therefore, it is interesting to numerically study large $N$ gauge
theory with fermions in the adjoint representation on a single
site lattice. Recent progress on this topic can be found 
in~\cite{Bedaque:2009md,Bringoltz:2009kb}.

\section*{Acknowledgments}

I wish to thank Jacek Wosiek, Michal Praszalowicz and other
organizers of the $49^{\rm th}$ Cracow School for their exceptional
hospitality
and a very lively summer school. I would like to acknowledge the
useful physics discussions with Jean-Paul Blaizot,  Mateusz Kore\'n, 
John Negele, Maciej Nowak,
Jan Pawlowski, Michael Teper and Jacek Wosiek
during the school. I am deeply indebted to my long term collaborator,
Herbert Neuberger, for involving me in several exciting problems
in particle physics. I would also like to acknowledge several physics
collaborations with Joe Kiskis that were covered in these
lectures.
This research was partially supported by the NSF under grant number
PHY-0854744.

\end{document}